\documentclass[letterpaper,aps,prl,twocolumn,superscriptaddress,amsmath, amssymb,showpacs]{revtex4}

\newif\ifpdf
\ifx\pdfoutput\undefined
    \pdffalse           
\else
    \pdfoutput=1        
    \pdftrue
\fi
\ifpdf
    \usepackage[pdftex]{graphicx}
    \usepackage{epstopdf}
\else
    \usepackage{graphicx}
\fi
\usepackage{dcolumn}
\usepackage{bm}
\usepackage{setspace}

\newcommand{\ppbar}{\ensuremath{\mathrm{p}\bar{\mathrm{p}}}}

\newcommand{\bbbar}{\ensuremath{\rm b\overline{b}}}
\newcommand{\ttbar}{\ensuremath{\rm t\overline{t}}}
\newcommand{\MET}{\mbox{$\raisebox{.3ex}{$\not$}E_{\rm T}$\hspace*{0.5ex}}}

\newcommand{\Pt}{\ensuremath{p_{\rm T}}}
\newcommand{\invfb}{\ensuremath{\rm fb^{-1}}\,}

\def\simle{\mathrel{
    \rlap{\raise 0.511ex \hbox{$<$}}{\lower 0.511ex \hbox{$\sim$}}}}
\def\etal{{\it et al.}}

\begin{document}

\hspace{5.2in} \mbox{FERMILAB-PUB-07-053-E}

\title{
  Inclusive Search for New Physics with Like-Sign Dilepton Events in
  \ppbar~Collisions at $\sqrt{s} = 1.96$~TeV
}
\affiliation{Institute of Physics, Academia Sinica, Taipei, Taiwan 11529, Republic of China} 
\affiliation{Argonne National Laboratory, Argonne, Illinois 60439} 
\affiliation{Institut de Fisica d'Altes Energies, Universitat Autonoma de Barcelona, E-08193, Bellaterra (Barcelona), Spain} 
\affiliation{Baylor University, Waco, Texas  76798} 
\affiliation{Istituto Nazionale di Fisica Nucleare, University of Bologna, I-40127 Bologna, Italy} 
\affiliation{Brandeis University, Waltham, Massachusetts 02254} 
\affiliation{University of California, Davis, Davis, California  95616} 
\affiliation{University of California, Los Angeles, Los Angeles, California  90024} 
\affiliation{University of California, San Diego, La Jolla, California  92093} 
\affiliation{University of California, Santa Barbara, Santa Barbara, California 93106} 
\affiliation{Instituto de Fisica de Cantabria, CSIC-University of Cantabria, 39005 Santander, Spain} 
\affiliation{Carnegie Mellon University, Pittsburgh, PA  15213} 
\affiliation{Enrico Fermi Institute, University of Chicago, Chicago, Illinois 60637} 
\affiliation{Comenius University, 842 48 Bratislava, Slovakia; Institute of Experimental Physics, 040 01 Kosice, Slovakia} 
\affiliation{Joint Institute for Nuclear Research, RU-141980 Dubna, Russia} 
\affiliation{Duke University, Durham, North Carolina  27708} 
\affiliation{Fermi National Accelerator Laboratory, Batavia, Illinois 60510} 
\affiliation{University of Florida, Gainesville, Florida  32611} 
\affiliation{Laboratori Nazionali di Frascati, Istituto Nazionale di Fisica Nucleare, I-00044 Frascati, Italy} 
\affiliation{University of Geneva, CH-1211 Geneva 4, Switzerland} 
\affiliation{Glasgow University, Glasgow G12 8QQ, United Kingdom} 
\affiliation{Harvard University, Cambridge, Massachusetts 02138} 
\affiliation{Division of High Energy Physics, Department of Physics, University of Helsinki and Helsinki Institute of Physics, FIN-00014, Helsinki, Finland} 
\affiliation{University of Illinois, Urbana, Illinois 61801} 
\affiliation{The Johns Hopkins University, Baltimore, Maryland 21218} 
\affiliation{Institut f\"{u}r Experimentelle Kernphysik, Universit\"{a}t Karlsruhe, 76128 Karlsruhe, Germany} 
\affiliation{High Energy Accelerator Research Organization (KEK), Tsukuba, Ibaraki 305, Japan} 
\affiliation{Center for High Energy Physics: Kyungpook National University, Taegu 702-701, Korea; Seoul National University, Seoul 151-742, Korea; and SungKyunKwan University, Suwon 440-746, Korea} 
\affiliation{Ernest Orlando Lawrence Berkeley National Laboratory, Berkeley, California 94720} 
\affiliation{University of Liverpool, Liverpool L69 7ZE, United Kingdom} 
\affiliation{University College London, London WC1E 6BT, United Kingdom} 
\affiliation{Centro de Investigaciones Energeticas Medioambientales y Tecnologicas, E-28040 Madrid, Spain} 
\affiliation{Massachusetts Institute of Technology, Cambridge, Massachusetts  02139} 
\affiliation{Institute of Particle Physics: McGill University, Montr\'{e}al, Canada H3A~2T8; and University of Toronto, Toronto, Canada M5S~1A7} 
\affiliation{University of Michigan, Ann Arbor, Michigan 48109} 
\affiliation{Michigan State University, East Lansing, Michigan  48824} 
\affiliation{Institution for Theoretical and Experimental Physics, ITEP, Moscow 117259, Russia} 
\affiliation{University of New Mexico, Albuquerque, New Mexico 87131} 
\affiliation{Northwestern University, Evanston, Illinois  60208} 
\affiliation{The Ohio State University, Columbus, Ohio  43210} 
\affiliation{Okayama University, Okayama 700-8530, Japan} 
\affiliation{Osaka City University, Osaka 588, Japan} 
\affiliation{University of Oxford, Oxford OX1 3RH, United Kingdom} 
\affiliation{University of Padova, Istituto Nazionale di Fisica Nucleare, Sezione di Padova-Trento, I-35131 Padova, Italy} 
\affiliation{LPNHE, Universite Pierre et Marie Curie/IN2P3-CNRS, UMR7585, Paris, F-75252 France} 
\affiliation{University of Pennsylvania, Philadelphia, Pennsylvania 19104} 
\affiliation{Istituto Nazionale di Fisica Nucleare Pisa, Universities of Pisa, Siena and Scuola Normale Superiore, I-56127 Pisa, Italy} 
\affiliation{University of Pittsburgh, Pittsburgh, Pennsylvania 15260} 
\affiliation{Purdue University, West Lafayette, Indiana 47907} 
\affiliation{University of Rochester, Rochester, New York 14627} 
\affiliation{The Rockefeller University, New York, New York 10021} 
\affiliation{Istituto Nazionale di Fisica Nucleare, Sezione di Roma 1, University of Rome ``La Sapienza,'' I-00185 Roma, Italy} 
\affiliation{Rutgers University, Piscataway, New Jersey 08855} 
\affiliation{Texas A\&M University, College Station, Texas 77843} 
\affiliation{Istituto Nazionale di Fisica Nucleare, University of Trieste/\ Udine, Italy} 
\affiliation{University of Tsukuba, Tsukuba, Ibaraki 305, Japan} 
\affiliation{Tufts University, Medford, Massachusetts 02155} 
\affiliation{Waseda University, Tokyo 169, Japan} 
\affiliation{Wayne State University, Detroit, Michigan  48201} 
\affiliation{University of Wisconsin, Madison, Wisconsin 53706} 
\affiliation{Yale University, New Haven, Connecticut 06520} 
\author{A.~Abulencia}
\affiliation{University of Illinois, Urbana, Illinois 61801}
\author{J.~Adelman}
\affiliation{Enrico Fermi Institute, University of Chicago, Chicago, Illinois 60637}
\author{T.~Affolder}
\affiliation{University of California, Santa Barbara, Santa Barbara, California 93106}
\author{T.~Akimoto}
\affiliation{University of Tsukuba, Tsukuba, Ibaraki 305, Japan}
\author{M.G.~Albrow}
\affiliation{Fermi National Accelerator Laboratory, Batavia, Illinois 60510}
\author{D.~Ambrose}
\affiliation{Fermi National Accelerator Laboratory, Batavia, Illinois 60510}
\author{S.~Amerio}
\affiliation{University of Padova, Istituto Nazionale di Fisica Nucleare, Sezione di Padova-Trento, I-35131 Padova, Italy}
\author{D.~Amidei}
\affiliation{University of Michigan, Ann Arbor, Michigan 48109}
\author{A.~Anastassov}
\affiliation{Rutgers University, Piscataway, New Jersey 08855}
\author{K.~Anikeev}
\affiliation{Fermi National Accelerator Laboratory, Batavia, Illinois 60510}
% ...
\collaboration{CDF Collaboration\footnote{With visitors from $^a$University of Athens, 
$^b$University of Bristol, 
$^c$University Libre de Bruxelles, 
$^d$Cornell University, 
$^e$University of Cyprus, 
$^f$University of Dublin, 
$^g$University of Edinburgh, 
$^h$University of Heidelberg, 
$^i$Universidad Iberoamericana, 
$^j$University of Manchester, 
$^k$Nagasaki Institute of Applied Science, 
$^l$University de Oviedo, 
$^m$University of London, Queen Mary and Westfield College, 
$^n$Texas Tech University, 
$^o$IFIC(CSIC-Universitat de Valencia), 
}}
\noaffiliation




\date{April 26, 2007}
\begin{abstract}
  We describe a search for anomalous production of events with two
  leptons ($e$ or $\mu$) of the same electric charge in
  \ppbar~collisions at a center-of-mass energy of $1.96$~TeV. Many
  extensions to the standard model predict the production of two
  leptons of the same electric charge.  This search has a significant
  increase in sensitivity compared to earlier searches. Using a data
  sample corresponding to 1~\invfb of integrated luminosity recorded
  by the CDF II detector, we observe no significant excess in an
  inclusive selection (expect $33.2\pm4.7$ events, observe $44$) or in
  a SUSY-optimized selection (expect $7.8\pm1.1$ events, observe
  $13$.)

\end{abstract}

\maketitle


The standard model of particle physics (SM) successfully describes all
experimental data taken in high energy collisions so far. Despite its
successes, there are strong indications that this theory is only an
effective low-energy model and new physics must be present at a higher
energy scale. An excellent signature to search for deviations from the
SM is production of two leptons with both leptons of the same electric
charge. This signature occurs naturally in many extensions to the SM
and occurs rather rarely in SM interactions.

An example of a model predicting like-sign dileptons is one with a
Majorana particle that decays through SM-like bosons into
leptons. Heavy Majorana neutrinos ($\nu_\mathrm{M}$) can be produced
in \ppbar~collisions in association with a lepton through a virtual
$\rm W$ boson ($\rm \ppbar\rightarrow\nu_M \ell^\pm X
$)~\cite{taohan}. This new particle can subsequently decay to a $\rm
W$ and another lepton ($\rm \nu_M\rightarrow W^\pm \ell^{\mp}$). Given
the Majorana nature of this neutrino, \emph{i.e.}, that it is its own
antiparticle, more than half of such events will contain like-sign
dileptons in the final state.
Another example is the class of models that predict new heavy analogs
to the $\rm W$ and $\rm Z$ bosons. For instance, in supersymmetric
extensions of the SM, a chargino-neutralino pair can be produced ($\rm
\ppbar\rightarrow \tilde\chi_1^\pm \tilde\chi_2^0X$) and decay into
final states with three charged leptons ($\rm
\tilde\chi_1^\pm\rightarrow \ell^\pm\nu \tilde\chi_1^0$ and $\rm
\tilde\chi_2^0\rightarrow \ell^\pm\ell^\mp
\tilde\chi_1^0$)~\cite{general_susy}. Two of those three leptons will
have the same charge.

The CDF and D\O\, Collaborations have previously investigated events
with two same-charge leptons~\cite{run1_lsdil,d0_lsdil}.  In this
Letter, we present a more general search using data collected with the
CDF~II detector during the Tevatron's Run II data-taking phase at a
center-of-mass energy of $1.96$~TeV.  We select events as inclusively
as possible without optimizing for any particular new physics
scenario. To avoid bias, we fix the final event selection criteria
before examining the event yield in the signal
region. 
The selection produces a relatively small sample that we investigate
for deviations from SM predictions, both in the total number of events
and in the shape of kinematic distributions.  We use a data sample
corresponding to 1~fb$^{-1}$ of integrated luminosity collected
between March 2002 and February 2006.  This search has better
acceptance and examines between a factor of three and a factor of ten
more integrated luminosity compared to earlier searches in the same
channel, resulting in roughly a factor of three increase in the
sensitivity to new physics.

The CDF~II detector is a general purpose particle detector and is
described in detail elsewhere~\cite{cdf2det}. It has a solenoidal
charged particle spectrometer, consisting of 7-8 layers of silicon
microstrip detectors and a cylindrical drift chamber immersed in a
1.4~T solenoidal magnetic field, a segmented sampling calorimeter, and
a set of charged particle detectors outside the calorimeter used to
identify muon candidates.  The fiducial region of the silicon
microstrip detector extends to $|\eta|\sim 2$~\cite{etpt}, while the
drift chamber provides tracking for $|\eta|\simle 1$. The curvature
resolution of the chamber is
$\sigma_C=3.6\times10^{-6}$\,cm$^{-1}$~\cite{cdf_wz_prd}. The
curvature corresponding to a track with momentum of 100\,GeV$/c$ is
$2.1\times10^{-5}$\,cm$^{-1}$.  The sign of the curvature of a track
with 100~GeV$/c$ of transverse momentum, and hence the charge of such
a particle, is thus typically determined with a significance of better
than five standard deviations.

We use data collected with a high-momentum central lepton trigger,
which identifies events with an electron candidate with $E_{\rm
  T}>18$~GeV and $|\eta|\simle1$ or a muon candidate with $p_{\rm
  T}>18$~GeV$/c$ and similar $\eta$ requirements.  We select events
with a pair of same-charge leptons (electrons or muons) regardless of
other activity in the event.  This analysis uses lepton candidates
with $|\eta|\simle1$. The tracks associated with the leptons have to
share a common vertex, \emph{i.e.}, they come from the same
\ppbar~interaction.  We define the two highest-momentum charged
leptons passing our selections as the leading and sub-leading lepton.
We select  leading and sub-leading electrons that fulfill the
following requirements on the transverse component of the energy:
$E_{\rm T}>20$~GeV and $E_{\rm T}>10$~GeV, respectively. We select
muons that pass similar transverse momentum requirements. We remove
photon-conversion electrons using a procedure described below.
Cosmic-ray muons are identified and removed by looking for a track
opposite to the reconstructed muon candidate which has timing
information that is consistent with a particle moving toward the
beam-line rather than away from it.  We also place a minimum
requirement on the invariant mass of the lepton pair,
$m_{\ell\ell}>25$~GeV/$c^2$, to remove the large background from
Drell-Yan and heavy quark production at low mass.  The leptons must be
isolated from other particles in the event, both in the calorimeter
and in the tracking chamber. An electron (muon) is considered to be
isolated in the calorimeter if the sum of the transverse energy within
a cone $\Delta R\equiv\sqrt{(\Delta\eta)^2 + (\Delta\phi)^2} \le 0.4$,
minus the lepton $E_{\rm T}$, is less than 10\% of the lepton $E_{\rm
  T}(p_{\rm T})$. Similarly, if the total transverse momentum of all
other tracks within a cone $\Delta R \le 0.4$ around the lepton is
less than 10$\%$ of the candidate track \Pt, the lepton is considered
to be isolated in the tracking chamber.

SM backgrounds that produce like-sign dileptons in the final state
include Drell-Yan dilepton production with a photon that is radiated
off a final-state lepton and converts into an
${\mathrm{e}}^+{\mathrm{e}}^-$ pair which fails our conversion
identification algorithm (\emph{i.e.}, is not ``tagged'') as well as
diboson production. The latter includes on- and off-shell $\rm ZZ$ and
$\rm WZ$ production followed by decay into leptonic final states, as
well as $\rm Z\gamma$ and $\rm W\gamma$ with the photon converting
into two electrons inside the detector. In the following, we refer to
the above Drell-Yan and $\rm Z\gamma$ processes as $\rm \ell \ell
\gamma$ backgrounds and to the $\rm W\gamma$ and $\rm WZ$ processes as
$\rm WV$.  Events with $\rm W(\to \ell \nu)+1$~jet and $\rm Z(\to
\ell\ell)+1$~jet where the jet is falsely identified as a lepton can
also result in two lepton candidates with the same
charge. Contributions from 
\ttbar, \bbbar\, backgrounds
are found to be negligible.

To estimate the background contribution from the $\rm\ell\ell\gamma$
and diboson processes, we determine geometric and kinematic acceptance
using Monte Carlo calculations followed by a {\sc geant}-based
simulation of the CDF II detector~\cite{geant3}.  We use the Monte
Carlo generator described in Ref.~\cite{wgamma} for the $\rm W\gamma$
background, {\sc madevent} for the $\rm WZ$
background~\cite{madgraph}, and {\sc pythia} for the other SM
processes~\cite{pythia}. We use the CTEQ5L parton distribution
functions to model the momentum distribution of the initial-state
partons~\cite{cteq6m}.  The expected number of events for each
background component is determined as the product of the cross
section, the luminosity of the sample, and the acceptance of the
detector. The last is corrected for trigger efficiency and differences
in lepton reconstruction efficiency between the data and the
simulation. These efficiencies are derived via studies of $\rm
W(\rightarrow \ell\nu$) and $\rm Z(\rightarrow \ell^+\ell^-$) events, and
the differences are typically less than 10\%.

Events with untagged photon conversions represent the dominant
background to this search. To tag photon conversions, we take
advantage of the kinematic condition that the trajectories of the
electron and the positron from the photon are approximately parallel
at the conversion point.  We define the following variables: $\Delta
s$ is the distance in the transverse plane between the two tracks at
the point that the two tracks are parallel, $\Delta\cot(\theta)$ is
the difference in the cotangents of the polar angles between the two
tracks, and $\Delta z$ is the distance in the $z$ dimension between
the two points on the tracks used to compute $\Delta s$.  A track pair
is tagged as a photon conversion if $\Delta s < 1.1$\,cm,
$|\Delta\cot(\theta)| < 0.26$ and $\Delta z < 1.2$\,cm.  These
thresholds are chosen to maximize the rejection power for conversions
while having a negligible impact on the efficiency for non-conversion
electrons.  Conversion tagging will fail due to inefficiencies of the
above selection or due to very asymmetric conversions, where the
momentum of either the electron or the positron drops below our track
selection threshold of $\Pt>500$~MeV/$c$.

Processes producing opposite-charge leptons, such as Drell-Yan
dilepton production, can contribute to our backgrounds if one of the
lepton tracks is poorly measured and its charge is incorrectly
reconstructed.  The charge of a particle is determined from the
direction the particle curves in the magnetic field. We test our
understanding of the charge misassignment mechanism by examining the
modeling of the curvature uncertainty. The uncertainty on the
curvature measured in the tracking chamber provides an estimate of the
probability of incorrectly measuring the sign of a track's curvature.
For this purpose, we select electrons from events with two electrons
where one electron is identified based solely on its energy deposits
in the calorimeter and examine the curvature uncertainty for this
particle. We find that the uncertainty on the track curvature of this
data sample is well described by our simulation of the detector. The
estimated residual background from events containing a lepton with an
incorrectly reconstructed charge is less than 0.1 events in the
current sample.

Jets in $\rm W$+jet and $\rm Z$+jet events can be misidentified as
leptons (``fakes'') and paired with the lepton from the gauge boson
decay to form a same-charge candidate event. We estimate this
background by selecting a sample of events with one or more
high-momentum leptons that pass our selection criteria, omitting
events with same-sign lepton candidates.  We determine the number of
events in our final selection from this fake background by multiplying
the number of isolated tracks in this sample by the misidentification
probability (``fake rate''). This probability is measured in a sample
triggered by at least one jet with $E_{\rm T}>50$~GeV and is defined
as the number of identified leptons, divided by the total number of
isolated tracks. We parameterize the fake rate as a function of
\Pt~and $\eta$.

As a further means of controlling untagged photon conversions, we
divide the electron candidates into two categories: with and without
energy depositions in the silicon microstrip detector (``silicon
hits''). Since most photon conversions occur either in the material of
this detector or in the inner wall of the drift chamber, electrons
with silicon hits are less likely to come from a conversion process.
We consider the more pure category of electrons with silicon hits
($\rm e_\mathrm{si}$) separately from those without ($\rm e$).  By
considering these two classes independently rather than requiring all
electrons to have silicon hits, we do not lose acceptance due to
inefficiencies in the silicon microstrip detector but gain in
statistical power. 

We perform numerous tests to assure that we are able to model our
backgrounds. These tests can be split into two categories: those that
test the overall normalization of a background, and those that test
our ability to model detector performance.  To probe the overall
normalization of the diboson $\rm Z\gamma$ background estimate, we
select events with two leptons and a photon with transverse momentum
thresholds of $20, 10$, and $10$~GeV$/c$, respectively. The predicted
number of events for this selection is $258\pm16$ events, dominated by
$\rm Z\gamma$ production.  We observe 258 events, in good
agreement. Similarly, we probe the normalization of the $\rm W\gamma$
background by selecting events with one lepton with $\Pt>20$~GeV$/c$,
a photon with $E_{\rm T}>10$~GeV, and $\MET>15$~GeV.  Using this
selection, we expect $1493\pm90$ events, dominated by $\rm W\gamma$
production. We observe 1540 events, in good agreement with our
expectation. For both of the above measurements, we require the photon
to be well-separated from either the electron or muon, thereby
effectively limiting the contribution from photons radiated in the
material of the detector. We check our modeling of the material in the
detector by selecting events with one electron, one photon, and
$\MET<20$~GeV. These are mostly Drell-Yan events in which one electron
has lost most of its energy to a radiated photon. For this selection,
we predict $243\pm15$ events and observe 269, thereby validating
our understanding of the detector material.  Other backgrounds are
tested using dedicated selection criteria. We obtain good agreement
between the observed and predicted events both in integral counts and
kinematic distributions in all regions considered.

\begin{table}[b]
\centering
\caption{Event counts 
  predicted
  ($n_\mathrm{pred}$) and observed ($n_\mathrm{obs}$), per
  category. The uncertainties for different channels are correlated.
  $n_\mathrm{\ell \ell \gamma}$, $n_\mathrm{WV}$, $n_\mathrm{ZZ}$, and
  $n_\mathrm{fake}$ refer to the number of background events predicted
  in the $\ell\ell\gamma$, $\rm WV(V=Z$ or $\gamma)$, $\rm ZZ$, and
  ``fake'' categories, respectively, and $n_\mathrm{pred}$ is the sum.
  $\rm e_{\rm si}$ and $\rm e$ refer to
  electron candidates with and without energy deposits in the silicon
  microstrip detector.  }
\label{tab:results:noZ:noMet}
  \begin{tabular}{l@{}rrrrrr}
\hline
    & $n_\mathrm{obs}$& ~~~~~~$n_\mathrm{pred}$&  $n_\mathrm{\ell \ell \gamma}$&
    $n_\mathrm{WV}$& $n_\mathrm{ZZ}$& $n_\mathrm{fake}$ \\ 
\hline
\hline
$\rm e_\mathrm{si}e_\mathrm{si}$ 
  & 11&   6.3$\pm$1.0&   3.2&   1.4& 0.4&    1.3 \\ 
$\rm ee$ 
  & 3&   1.3$\pm$0.3&   0.9&   0.1&   0.0&   0.2 \\ 
$\rm e_\mathrm{si}e$ 
  & 9&   9.1$\pm$1.8&   6.4&   1.6&   0.1&    1.0 \\ 
$\rm e_\mathrm{si}\mu$ 
  & 11&   6.8$\pm$0.8&   0.8&   2.8&   1.1&  2.1\\ 
$\rm e\mu$ 
  & 5&   6.4$\pm$1.2&   3.4&   1.9&   0.2&   0.9 \\ 
$\rm \mu\mu$ 
  & 5&   3.2$\pm$0.3&   0.1&   1.4&   0.8&     0.8 \\ 
\hline
Total 
 & 44&  33.2$\pm$4.7&  14.9&   9.3&   2.5&     6.4 \\ 
\hline
\end{tabular}
\end{table}
The uncertainty on the number of predicted background events is
dominated by the uncertainty on the luminosity measurement ($5\%$),
the fake estimate ($5\%$) and the conversion modeling ($10\%$). Other
uncertainties include those associated with the SM cross sections,
lepton reconstruction, and the statistical uncertainty on the Monte
Carlo acceptance calculation.

New physics scenarios that lead to like-sign dilepton events, such as
the ones mentioned in the introduction to this Letter, often have a
$\rm WZ$-like topology.  As such, SM $\rm WZ$ production provides a
reference point for the sensitivity of this analysis.  The product of
geometric acceptance, kinematic acceptance, and like-sign dilepton
identification efficiencies (${\cal A_{\rm geo} \times A_{\rm
    kin}}\times \epsilon$) for leptonic decays for on-shell SM $\rm 
WZ$ production, given as a proxy for the sensitivity of this analysis,
is $8.0\%$.

\begin{figure*}[htbp]
\begin{center}
  \caption{Invariant mass distribution $m_{\ell\ell}$ of the 
    selected leptons (a), $\not$${E_T}$ (b), leading (c)
    and sub-leading (d) lepton transverse momentum in data and
    simulation.  The rightmost bins are overflow bins. The peak at
    $m_{\ell\ell}\approx 90$ GeV is mostly due to Drell-Yan
    di-electron production with  hard radiation off one electron,
    followed by an asymmetric conversion.  }
 \label{fig:signal}
 \includegraphics[width=0.9\textwidth]{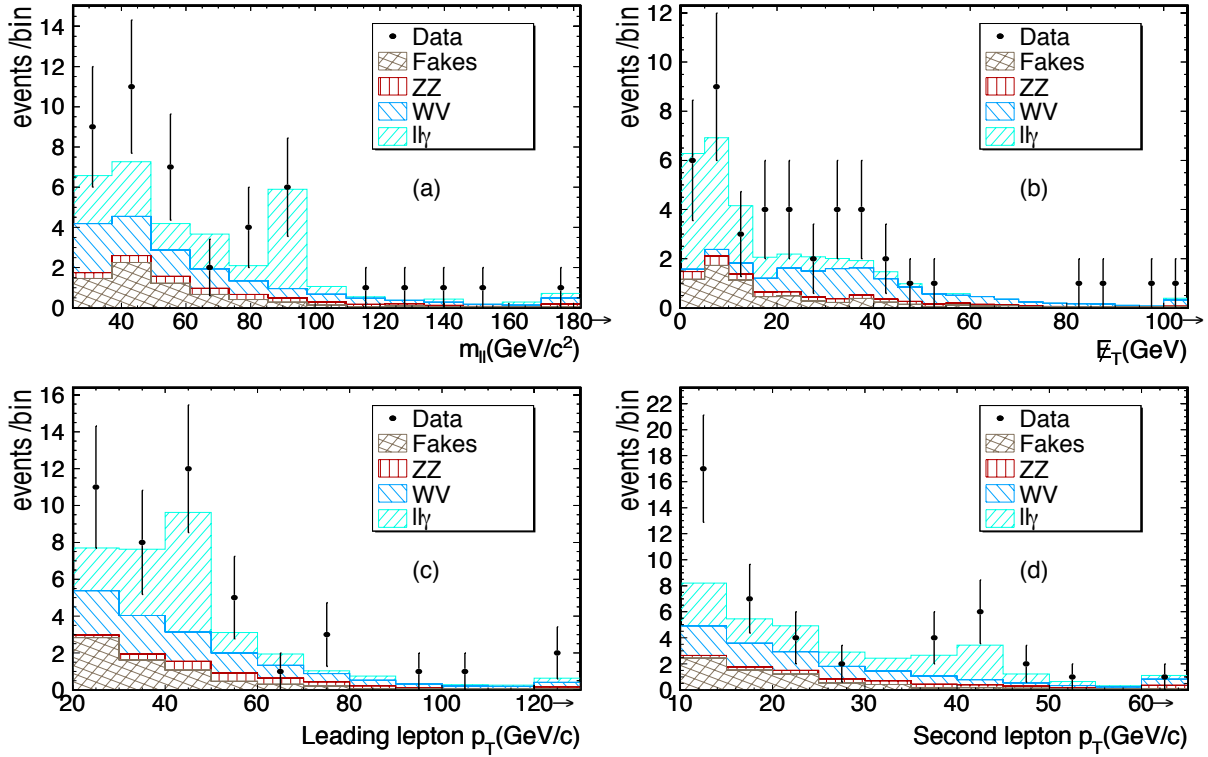}
\end{center}
\end{figure*}

In Table~\ref{tab:results:noZ:noMet}, we present the expected and
observed number of events with two same-charge leptons in all
different combinations of leptons considered.  
Combining all channels, we predict
$33.2\pm4.7$ events and observe $44$.  The probability that $33.2$
events with an uncertainty of 4.7 events fluctuate to $44$ or more is
$9\%$.  
Figure~\ref{fig:signal} shows the comparison between the predicted and
observed events for several kinematic distributions of interest.
\begin{table}
  \centering
  \caption{Event counts 
    with additional selection criteria for a SUSY-like physics
    scenario. The table headings are explained in the caption to
    Table~\ref{tab:results:noZ:noMet}.
  }
  \label{tab:results:z:met}
  \begin{tabular}{lrrrrrr}
    \hline
    & $n_\mathrm{obs}$& ~~~~~~$n_\mathrm{pred}$&  $n_\mathrm{\ell \ell \gamma}$&
    $n_\mathrm{WV}$& $n_\mathrm{ZZ}$& $n_\mathrm{fake}$\\ 
    \hline
    \hline
    $\rm e_\mathrm{si}e_\mathrm{si}$ 
    & 1&   1.3$\pm$0.3&   0.4&   0.6&   0.0 & 0.4 \\
    $\rm ee$ 
    & 1&   0.1$\pm$0.1&   0.0&   0.1&   0.0 & 0.0 \\
    $\rm e_\mathrm{si}e$ 
    & 2&   1.5$\pm$0.3&   0.1&   1.2&   0.0 & 0.2 \\
    $\rm e_\mathrm{si}\mu$ 
    & 4&   1.7$\pm$0.2&   0.0&   1.0&   0.1 & 0.7 \\
    $\rm e\mu$ 
    & 4&   2.3$\pm$0.5&   0.6&   1.4&   0.0&   0.2 \\ 
    $\rm \mu\mu$
    & 1&   0.9$\pm$0.1&   0.0&   0.5&   0.1&   0.4 \\ 
    \hline
    Total 
    & 13&  7.8$\pm$1.1&   1.1&   4.7&   0.2&   1.8 \\
    \hline
  \end{tabular}
\end{table}

In Table~\ref{tab:results:z:met}, we present the expected and observed
number of events after imposing two additional requirements aimed at
increasing the signal sensitivity for a SUSY-like physics scenario
where the stable, neutral and lightest supersymmetric particle escapes
detection.  We require a large transverse momentum imbalance
($\MET>15$~GeV) and reject events in which the invariant mass of one of
our selected leptons and another lepton of the same flavor and
opposite charge are consistent with a $\rm Z$ boson
($66<m_{\ell^+\ell^-}<116~\mathrm{GeV}/c^2$). We predict $7.8\pm1.1$
events and observe $13$.  The probability that $7.8$ events with an
uncertainty of 1.1 events fluctuate to $13$ or more is $7\%$.  The
value of ${\cal A_{\rm geo} \times A_{\rm kin}}\times \epsilon$ for
$\rm WZ$ events, as described above but adding a $\MET>15$~GeV
requirement, is 7\%.

In conclusion, we have performed a search for events with two leptons
of the same electric charge using the CDF Run II data. We observe a
slight excess in the number of predicted events in almost all lepton
categories.  However, the kinematic distributions do not show any
anomalous deviation from expectations in any particular region of
parameter space.  Large future data sets expected at the
Tevatron will reveal whether this observed slight excess persists.

We thank the Fermilab staff and the technical staffs of the
participating institutions for their vital contributions.  This work
was supported by the U.S. Department of Energy and National Science
Foundation; the Italian Istituto Nazionale di Fisica Nucleare; the
Ministry of Education, Culture, Sports, Science and Technology of
Japan; the Natural Sciences and Engineering Research Council of
Canada; the National Science Council of the Republic of China; the
Swiss National Science Foundation; the A.P. Sloan Foundation; the
Bundesministerium f\"ur Bildung und Forschung, Germany; the Korean
Science and Engineering Foundation and the Korean Research Foundation;
the Particle Physics and Astronomy Research Council and the Royal
Society, UK; the Russian Foundation for Basic Research; the Comision
Interministerial de Ciencia y Tecnologia, Spain; and in part by the
European Community's Human Potential Programme under contract
HPRN-CT-2002-00292, Probe for New Physics.



\begin{thebibliography}{99}
\bibitem{etpt} We use a cylindrical coordinate system about the beam
  axis in which $\theta$ is the polar angle, $\phi$ is the azimuthal
  angle, and $\eta \equiv -\ln \tan (\theta /2)$. $E_{\rm T} \equiv E
  \sin \theta$ and $p_{\rm T} \equiv p \sin \theta$ where $E$ is
  energy measured by the calorimeter and $p$ is momentum measured by
  the spectrometer. We define the missing transverse momentum
  $\vec{\MET}$$\equiv -\sum_i E_{\rm T}^i {\bf n}_i$, where ${\bf
    n}_i$ is the unit vector in the azimuthal plane that points from
  the beamline to the $i$th calorimeter tower.
\end{thebibliography}
\end{document}